\documentclass[12pt]{iopart}
\usepackage{epsf} 
\usepackage{graphics}
\usepackage{amssymb}
\usepackage{iopams}

\begin{document}

\title[Back-of-the-envelope climate model]{Validating an old physics-based back-of-the-envelope climate model with state-of-the-art data}

\author{Rasmus E. Benestad}

\address{The Norwegian Meteorological Institute}
\ead{rasmus.benestad@met.no}

\begin{abstract}
An old conceptual physics-based back-of-the-envelope model for greenhouse effect is revisited and validated against state-of-the-art reanalyses. Untraditional diagnostics show a physically consistent picture, for which the state of earth's climate is constrained by well-known physical principles, such as energy balance, flow and, conservation. Greenhouse gas concentrations affect the atmospheric optical depth for infrared radiation, and increased opacity implies higher altitude from which earth's equivalent bulk heat loss takes place without being re-absorbed. Such increase is seen in the reanalyses. There has also been a reduction in the correlation between the spatial structure of outgoing long-wave radiation and surface temperature, consistent with increasingly more processes interfering with the upwelling infrared light before it reaches the top of the atmosphere. State-of-the-art reanalyses further imply increases in the overturning in the troposphere, consistent with a constant and continuous vertical energy flow. The association between these aspects can be interpreted as an entanglement between greenhouse effect and the hydrological cycle, where reduced energy transfer associated with increased opacity is compensated by tropospheric overturning activity. 
\end{abstract}


\bibliographystyle{Science} 

\maketitle

\section{Motivation}


The essential principles of the greenhouse effect (GHE) have long been understood \cite{Weart2003} in terms simple reductionistic models \cite{Fleagle80,Houghton91,Peixoto92,Hartmann94,Lacis2010,Pierrehumbert2011}, however, these models usually don't provide a comprehensive description of the complexity associated with an anthropogeic global warming (AGW). Present global climate models supersede more simple and conceptual models on the greenhouse effect, such as radiative-convection and heat balance models discussed between 1900--1980 \cite{Arrhenius1896,Hulburt1931,Charney79,Schneider74,North75,Wang80}. State-of-the-art global climate models (GCMs), on the other hand, are too complex for grasping the basic elements. 
Yet, revisiting an old 'back-of-the-envelope', proposed by Hulburt in 1931 \cite{Hulburt1931}, can provide us with a greater appreciation for how increased CO$_2$-concentrations affect our atmosphere, and as proposed by Schneider and Dickinson in 1973 \cite{Schneider74}, provide some didactic value in addition to explaining the essence of the greenhouse effect. 
  
The didactic aspect is still relevant, as there is still a number scientists who doesn't acknowledge the link between increased levels of greenhouse gas concentrations (GHGs) and climate change \cite{Anderegg2010}, and who are not persuaded by the scientific literature supporting the notion of an anthropogenic global warming (AGW) \cite{Oreskes2004,IPCC-AR4}. 
Furthermore, there is a surprisingly high number of phycisists among this group \cite{OreskerConway2008} which could hint that old knowledge has been forgotten \cite{ProctorSchiebinger2008}.
By revisiting Hulburt's 1931 physics-based back-of-the-envelope climate model \cite{Hulburt1931} in the context of modern state-of-the-art data, the objective is to show that the model is consistent with data which were not available in 1931, including recent years as well satellite-based measurements of outgoing long-wave radiation and several reanalyses.

\subsection{A simplified physical picture}

The starting point for the model is that the energy flows through the universe, being generated inside the sun and intercepted by the planets \cite{Trenberth2004}. When the energy leaves the terrestrial system, it must be in the form of electromagnetic radiation, since space is a vacuum and does not provide a medium for other type of energy transfer \cite{Pierrehumbert2011}. 

The radiative energy emitted from an object can often be described by Planck's law, relating electromagnetic energy loss to the '{\em emission temperature}' ($T_e$). The terrestrial energy loss is often referred to as the {\em outgoing long-wave radiation} (OLR).  The total energy loss may be approximated by a bulk flux \cite{Hulburt1931}, although spectral lines due to atomic and molecular line absorption and quantum mechanics play an important role \cite{Fleagle80,Houghton91,Pierrehumbert2011}. The rate of heat loss must equal the rate of energy received from the sun for a planet to be in energetic equilibrium \cite{Hulburt1931}. The planetary energy balance can then be described approximately by the simple equation 

\begin{equation}
S_0 (1-A)/4 = \sigma T_e^4,
\label{eq:1}
\end{equation}

\noindent where $\sigma=5.67 \times 10^{-8}W/(m^2 K^4)$ is the Stefan-Boltzman constant, $A \sim 0.3$ is the albedo, and $S_0=1366W m^{-2}$ is the 'solar constant' \cite{Schneider74,Wang80}. The left hand side of this equation represents the energy input while the right hand side describes the bulk heat loss. Predicted temperatures according to equation \ref{eq:1} can be compared with actual observed surface temperatures in our solar system to demonstrate predictive skill\footnote{See Fig S2 in SM}. The value of $T_e$ derived from measured terrerstrial OLR is typically in the range 251--254K\footnote{Fig S3 in SM}.

Most of the energy from the sun is absorbed at the planet's surface, as earth's cloud-less atmosphere is transparent to visible light while clouds tend to reflect the light. Equation \ref{eq:1}, however, implies a value of $T_e=254K$ for the terrestrial emission temperature, whereas the observed global mean surface temperature is $T \approx 288K$. The bulk of the heat loss to space cannot take place at the ground level where $T=288K$ because this would violate the energy balance (equation \ref{eq:1}). However, it has long been established that the planetary heat loss is determined by temperature of the atmosphere far above the surface \cite{Hulburt1931,North75}. The temperature drops with height due to convective adjustment and the radiative heating profile, and equals the emission temperature of 254K at around 6.5 km above the ground\footnote{Fig S9 in SM}. This altitude is where earth's bulk heat loss must take place according to equation \ref{eq:1}. This is also independently corroborated by $T_e$ derived from measured OLR\footnote{See Fig S3 SM}.  

The GHE consists of the air being opaque to light in the long-wave range yet transparent for short-wave radiation. The degree of transparency of a medium is described by Beer's law, which relates the optical depth to the concentration of the absorbing medium and the medium's absorbing capacity \cite{Fleagle80,Peixoto92}. Several atmospheric gases (e.g. H$_2$O, CO$_2$ in addition to clouds) are opaque in dominant frequency ranges predicted by Plank's law for $T\approx 288K$. Indeed, the atmospheric CO$_2$-concentrations are specified from infrared gas analyser measuring the amount of infrared (IR) light absorbed in air samples \cite{Keeling76}. 

In the atmosphere, IR light is expected to be absorbed and re-emitted multiple times before its energy reaches the emission level where it is free to escape to space \cite{Pierrehumbert2011}. The process of absorption and re-emission will result in a more diffuse structure for the OLR at the top of the atmosphere. Hence, for an observer viewing the earth from above (e.g. a satellite instrument measuring the OLR), the bulk IR light source would be both more diffuse and located at increasing heights with greater GHGs, as the depth to which the observer can see into the atmosphere gets shallower for more opaque air. This altitude is henceforth referred to as the 'bulk emission level' $Z_{T254K}$, which represents the mean height for both cloudy and cloud-free regions. 

Higher concentrations of greenhouse gases divert more IR radiation downwards toward the ground \cite{Trenberth2011}. Hence, changes in the opacity alters the vertical energy flow, which needs to be compensated if the planet is to remain in energy balance and the total energy transfer continuous \cite{Hulburt1931}.
A deeper optical depth, due to increased absorption, is expected to act as 'resistance' for the radiative energy transfer, everything else being constant. Reduced radiative energy flux must be compensated though e.g. increased temperatures, or latent/sensible heat fluxes.

\section{Methods \& Data}

The height of the global mean 274K isotherm $Z_{T254K}$ was estimated from 37 model levels between 1000hPa and 1hPa (which corresponds to $\sim$0m and $\sim$63000m above sea level) from the European Centre for Medium-range Weather Forecasts (ECMWF) interim reanalysis (ERAINT) reanalysis \cite{ERAINT}. Monthly mean values for $T$ were estimated from instantaneous values, sampled at four times a day over the interval with available data Jan-1979--Sep-2012 at the full $0.75^\circ \times0.75^\circ$ horizontal resolution. 

The monthly mean vertical velocity $w$ from ERAINT was used provide a metric for the global atmospheric overturning. 
Convection is characterised by a circulation pattern of rising and sinking air masses. The atmospheric vertical volume transport takes place through cells of updraught and subsidence, however, these may not be coherent or stable in time and space. It was therefore important to find a metric that doesn't assume patterns that are stable in time. 

The global spatial variance in the vertical mass transport for a given moment in time was used to represent the state of vertical overturning on a global scale. Sensible and latent heat transfer are related to the vertical motion, but not completely determined by the flow as the atmospheric vapour content and temperature gradients also are important factors. However, water vapour content is poorly constrained, due to uncertainties associated with clouds and precipitation. 
The vertical velocity ($w$), on the other hand, is related to horizontal air flow through divergence, convergence, and variation in well-described quantities such the geopotential heights and barometric pressure. 

In addition to the ERAINT reanalysis, the NCEP/NCAR and ERA40 reanalyses were also used as these covered longer time intervals. OLR, measured from space \cite{LiebmannSmith2006}, was used to test equation 1 as well as analysing trend in opacity to IR radiation. Moreover, OLR measured from satellites \cite{LiebmannSmith2006} was used to derive $T_e$ and examine trends in diffuse IR emission at the top of the atmosphere. See the SM for more details and listings of scripts used for the analysis.

\section{Results}

The 254K isotherm $Z_{T254K}$ represents the equivalent altitude where earth's bulk heat emission takes place, and an upward trend of 23m/decade is consistent with a deeper optical depth as well as a global mean surface warming of 0.12K/decade over the period 1979--2011, not widely different to published warming rates of 0.14 to 0.18 K/decade \cite{Foster2011}. Figure 1 shows two different estimates of $Z_{T254K}$ based on the ERAINT and the NCEP/NCAR reanalysis (black and grey curves respectively). There was a good general agreement between the two, albeit with some differences in the detailes. This means that an observer in space would see IR radiation emanating from a shallower atmospheric depths as the GHGs have increased over time\cite{IPCC-AR4}. 

Figure 2 shows the time evolution of the pattern correlation between the OLR and the T(2m) from the ERA40 reanalysis. The correlation estimates were in general high due to the warm tropics and cool polar regions, which also influenced the local heat loss. There were also large short-term variations, which were likely to be affected by annual changes in the cloud-cover. Nevertheless, the estimates over the interval 1975--2002 suggested a decreasing trend which is consistent with the spatial structure of the OLR becoming more diffuse due to greater opacity, either due to increased GHE or changed cloudiness. High cloud tops, however, are consistent with high $Z_{T254K}$ and the simple model, whereas the OLR from low cloud tops are more similar to the upwelling IR from the surface.

Peaks associated with ENSO, seen in the $Z_{T254K}$, however, were less visible in the correlation, although there may be weak hints low correlation around some of the El Ni\~{n}o years. One explanation for weak ENSO sigature may be that the OLR is expeced to be sensitive to the cloud cover, which may dominate on the short time scales, and that ENSO involves longitudinal shifts rather than latitudinal changes\footnote{See the SM}.

Figure 3 shows $\eta_z(t)=\sum_z var(a_i w_i(z,t))$ for 3 different levels in the troposphere. The variability above $\sim$1000 m a.s.l has, according to ERAINT, increased since 1995, with most pronounced increase in the middle troposphere ($\sim$1km--6.5km a.s.l.; black). The upward trend in the middle atmosphere is consistent with the notion that increased convection compensates for reduced radiative transfer between the ground and $Z_{T254K}$. The upper layer above $\sim$6.5km a.s.l. also exhibited a smaller trend, but this layer may have been affected by the convective activity below. 

The blue curve in Figure 1 shows the vertical integral of total column water vapour ($q_{tot}$), exhibiting an insignificant trend. Hence, the increased overturning supports the notion of an enhanced vertical latent heat transport. The excursions in both $q_{tot}$, and $Z_{T254K}$ in 1988, 1991--1992, 1997--1998, and 2009-2010 may be associated with ENSO, and demonstrate how the optical depth is affected by changes in the atmospheric moisture or the clouds. 

The global variations in $\eta_z(t)$ in the lower 1000m, however, had an inter-annual time scale, but nevertheless exhibited no clear association with the El Ni\~{n}o Southern Oscillation. This layer embeds most of the planetary boundary layer (PBL), which is affected by the surface friction and turbulent mixing, whereas the air above represents the free atmosphere \cite{Schneider74}. In the lower PBL, the main energy transport takes place through small-scale turbulence with short time scales which may not be well represented by monthly mean values from the atmospheric model used for the reanalysis. Furthermore, the PBL $w$ may not have been assimilated well, due to scarce observations for that level and the incomplete understanding of this lowest layer of air. 

\section{Discussion}

The results here suggest that that a simple model provides an approximate description, by treating all the IR light as one bulk heat loss at the top of the atmosphere where it leaves the planet, as in equation \ref{eq:1}, and defines the bulk emission altitude as the region where the temperature equals the emission temperature $T=T_e$ \cite{Houghton91}. 
A similar treatment of the bulk emission and its vertical level is also the basis for the discussion on the 'saturation effect' in the report 'The Copenhagen Diagnosis' (2009)\cite{TheCopenhagenDiagnosis2009}. 

The trend analysis for the 254K isotherm $Z_{T254K}$ can be considered as an extension of Santer et al. (2003) \cite{Santer2003b}, who reported a robust, zero-order increase in tropopause height over 1979--1997 in two earlier versions of re-analyses, which they interpreted as an integrated response to anthropogenic forced warming of the troposphere and cooling of the stratosphere. Here, this aspect is put into a simple physics context of energy flow, where the metric $Z_{T254K}$ is interpreted as an equivalent mean level for the bulk emission. This type of representation is a simplification on par with 'model physics' such as parametrisation of clouds in the GCMs themselves. Such simple models are strictly not correct but may nevertheless be useful according to M.I. Budyko (1974): ``some aspects of climate genesis might be elucidated by means of the simplest models'' \cite{Schneider74}. 

Hence, the emission level diagnostic is not actually an emission level, it's just the height of the temperature that corresponds to the bulk emission level on the right hand side of equation \ref{eq:1}. The objective here was to provide a simple picture of an enhanced GHE and an 'educational toy' \cite{Schneider74} that is consistent with the most up-to-date analysis of the atmosphere.

The consideration of the vertical energy flow, where latent heat tranfer and convective processes compensate for reduced IR energy transfer, entangles the GHE with the hydrological cycle. Convection and latent heat both set the atmosphere's vertical temperature profile and involve evaporation, condensation, and precipitation. From these principles, an increased GHE is expected to imply changes in the rainfall patterns\footnote{Fig S7 in the SM}. 

It is interesting to note that the trend analysis in the atmospheric overturning metric presented here appears to be inconsistent with the previous conclusions \cite{Vecchi2007} suggesting that the strength of the atmospheric overturning circulation {\em decreases} as the climate warms. This apparent disparency is interesting and merits further investigation into whether it is caused by different metrics or whether the GCMs really provide a different description to the reanalyses.
On the other hand, if one considers the energy flow from the surface to the emission levels higher aloft, and that an increased optical depth diminishes the vertical energy transfer associated with IR radiation, a reduction in convective activity is surprising, as the flow of energy must be continuous and any divergence in the total flux will lead to deficit or surplus of heat. A simulated warming in the upper troposphere \cite[Figures 9.1 \& 9.2]{IPCC-AR4} nevertheless suggests an enhanced emission of IR from greater heights, which must be supported by an increased vertical energy flow. It is, however, possible that an increased latent heat flux, due to increased atmospheric moisture, can maintain this even if the overturning were to slow down. It is also conceivable that energy is transferred through other means of energy transfer, such as the propagation of gravity waves which tend to be represented by the means of parametrisation in atmospheric models. 
A change in $\eta(z,t)$ can also be linked to reports of changes to the Hadley cell \cite{Seidel2008,Giorgi-et-al-2011}, and changes in convective activity may have some impact on the hydrological cycle as these tend to involve cloud formation. Earth's heat loss is strongly tied to clouds, as cloud tops are prominent features in the OLR \cite{Wang2002}. 

The reanalyses provide the to-date best picture that we have of the atmosphere and an improvement over previous studies \cite{Santer2003b}, but there are some caveats as the introduction of new instruments, such as satellite instruments, will result in inhomogeneity\footnote{http://climatedataguide.ucar.edu/} \cite{Hines2000,Bengtsson2004}. It is especially the introduction of new satellite data assimilation in the reanalysis that cause non-climatic changes. Furthermore, unresolved processes, such as cumulus convection may not be well-represented, and monthly means, used for the vertical velocity, do not necessarily capture fluxes taking place on shorter timescales. Nevertheless, the old radiative-convection and heat balance models appeared to be consistent with the long-term trends found in the reanalyses. Increased atmospheric overturning and emission level altitude support the conceptual picture of an enhanced GHE, for which the energy flow between the ground and the emission level must be constant. The spatial correlation between the OLR and surface temperature provides an indication of the strength of GHE, but is also strongly affected by clouds. Clouds, however, are part of the GHE, which subsequently affect the bulk emission level $Z_{T254K}$.

Although the model presented here was similar to the ideas discussed by Hulburt in in 1931, he did not emphasis on the vertical energy flow, the hydrological cycle, and changes in the greenhouse gas concentrations. Although these aspects were were implicit in his model, he did not extend these to the question of climate change and the entanglement between the hydrological cycle and the GHE. An interesting question is whether reviving this old physics-model in the context of modern data can convince scientifically literate who still don't acknowledge an AGW. There is a cause for optimism if the reason for different views is 'agnotological' in terms of forgotten ideas \cite{ProctorSchiebinger2008}.

\section{Conclusions}

Physical constraints provided by the simple conceptual model introduced by Hulburt in 1931 \cite{Hulburt1931} provide untraditional diagnostics of an enhanced greenhouse effect in reanalyses and the OLR. These reanalyses, however, may involve serious caveats in terms of time-dependent errors and biases, and the analysis has not ascertained whether these trends are real. The long-term trends identified are nevertheless consistent with the noition of increasing opacity for IR light, elevation of the OLR emission level, and convective activity. This framework embeds more sophisticated explanations than traditional purely radiative models, and incorporate effects from feedback processes through $dT/dz$, response in $Z_{T254K}$, and the albedo $A$. It further provides a link to the hydrological cycle.

\section*{Figures and figure captions}

\clearpage
\newpage

\begin{figure}
\includegraphics{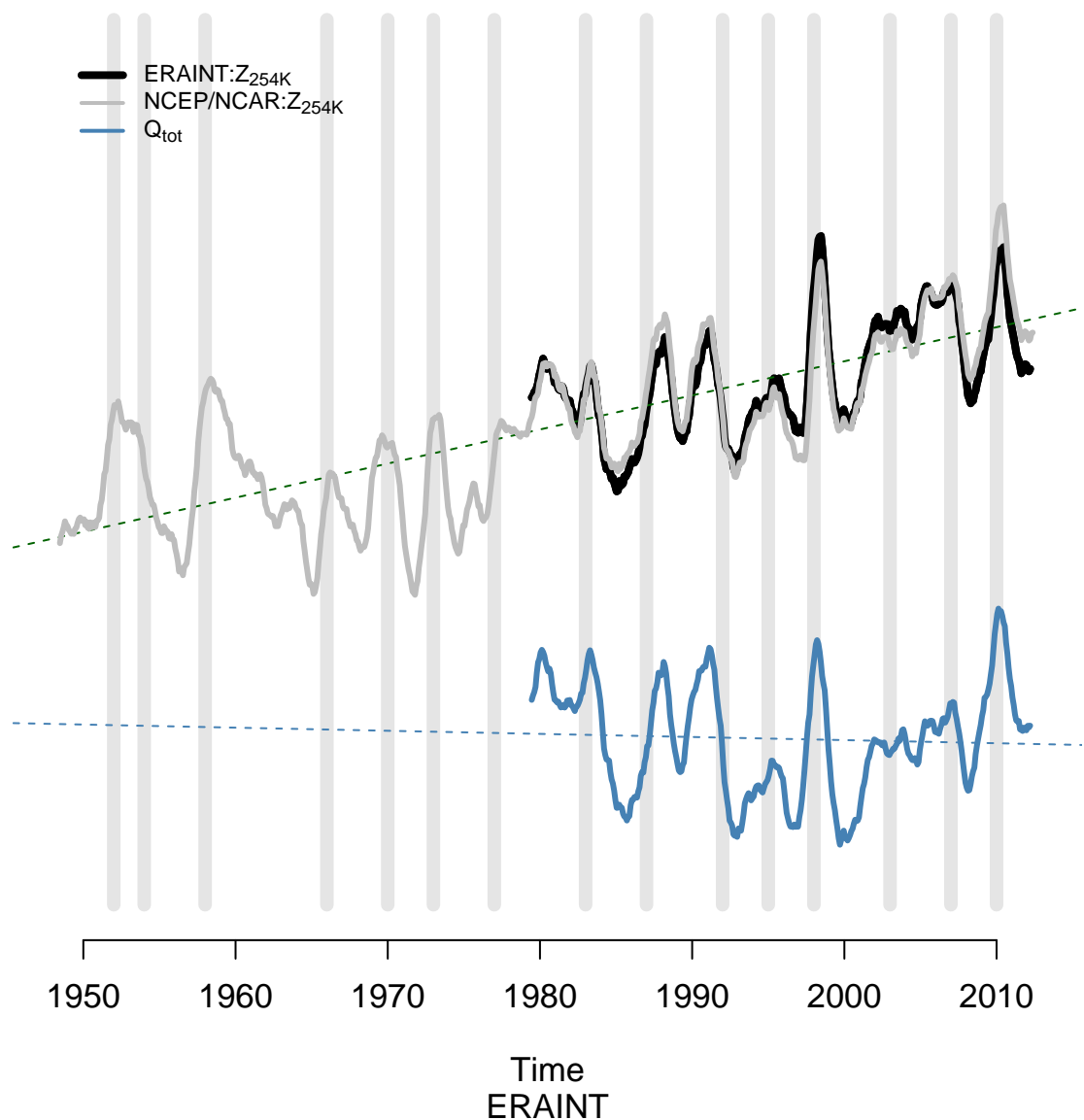}
\caption{\label{fig:2} The 12-month moving mean of the spatial variance of the vertical volume transport the vertical integral of total column water vapour ($q_{tot}$; blue), and the bulk emission level altitude ($Z_{T254K}$; black) from ERAINT. The mean $Z_{T254K}$ was estimated to be $7219 \pm 2$m, and the trend in the altitude of the bulk emission level, $(23 \pm 2) m/decade$, supports the notion of increased optical depth and hence and enhanced GHE. The trend in $Q_{tot}$ was $(-0.018 \pm 0.017) kg/(m^2 decade)$ (mean=$29kg/m^2$). The curves are plotted here with arbitrary scales along the y-axis. The grey vertical bands mark January 1st of known El Ni\~{n}o years.}
\end{figure}

\clearpage
\newpage

\begin{figure}
\includegraphics{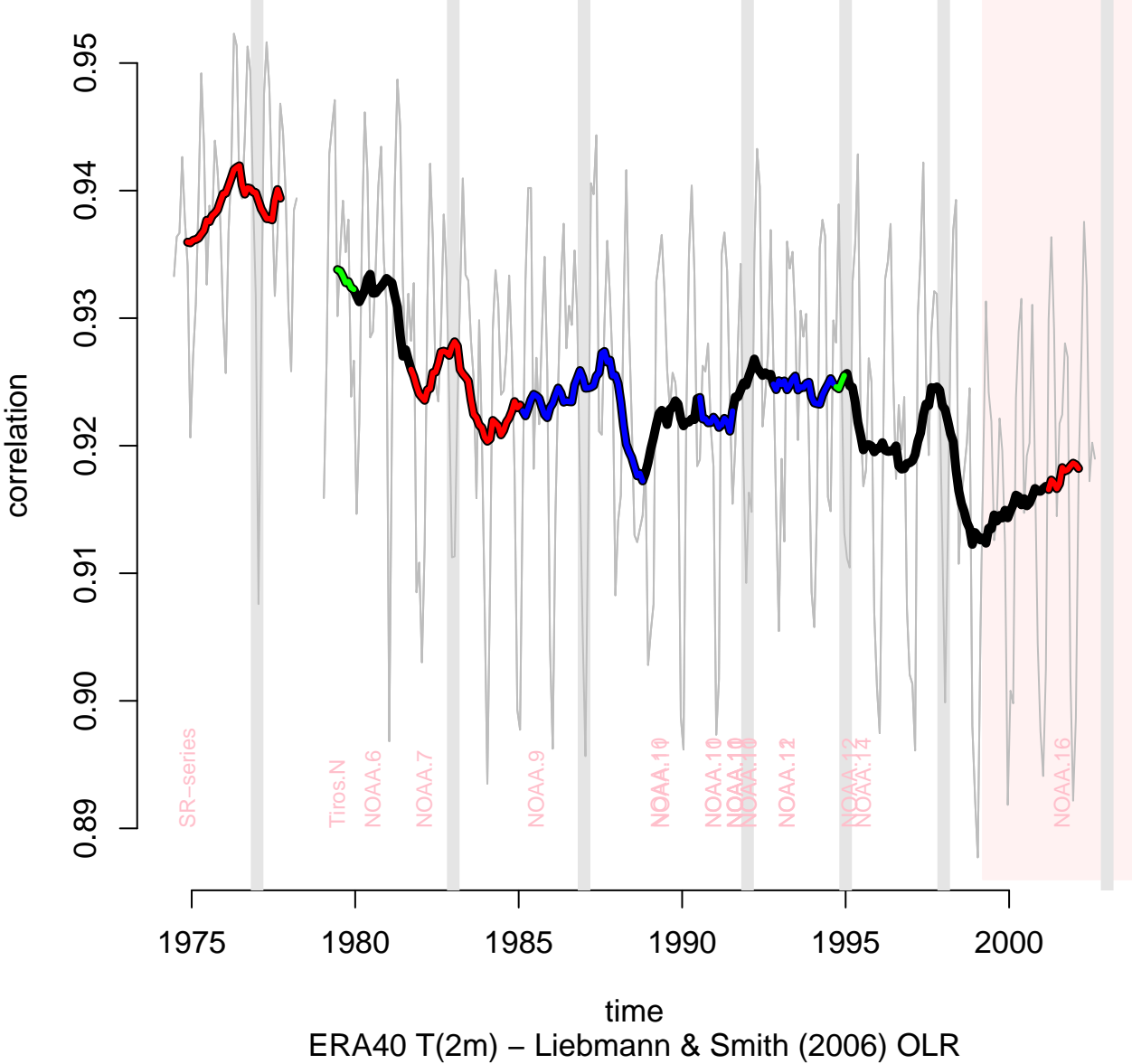}
\caption{\label{fig:3} The spatial correlation between the OLR and T(2m) estimated for each month over the period 1975--2002. The grey curve is unfiltered results and the thick black shows a 12-month moving average. The long-term reduction is consistent with a more blurred OLR pattern compared to the surface temperature. The different satellite missions are shown in different colours for the low-pass filtered curve. The faint pink region marks new interpolation scheme.}
\end{figure}

\clearpage
\newpage

\begin{figure}
\includegraphics{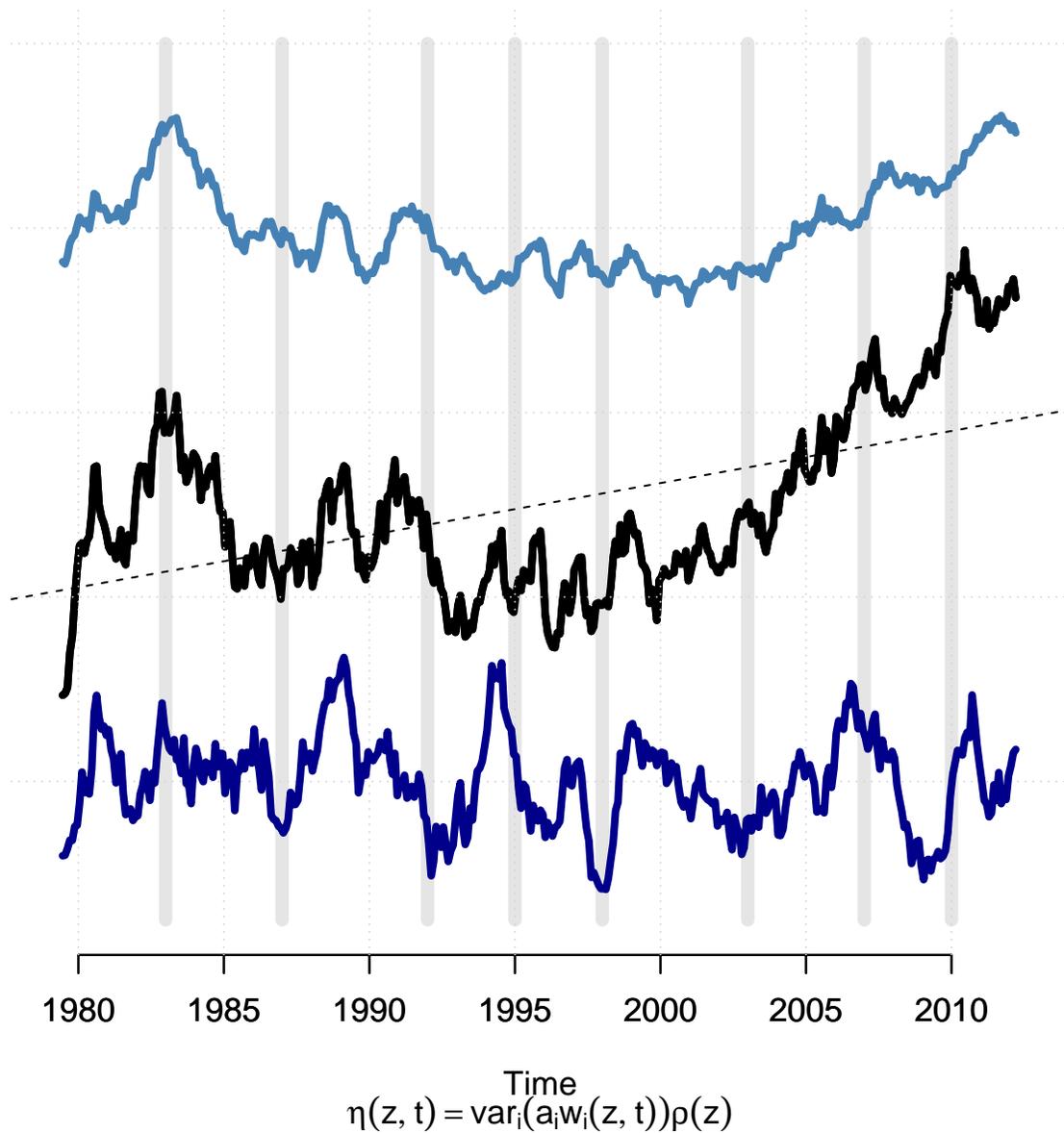}
\caption{\label{fig:1} The 12-month moving mean of the spatial variance of the vertical volume transport ($var(a \times w)$) from ERAINT.  The trend in the atmospheric overturning, $(1100 \pm 90) kg s^{-1}/decade$, in the atmospheric middle levels (black) supports the notion of increased optical depth and hence and enhanced GHE. The curves are plotted here with arbitrary scales along the y-axis.}
\end{figure}

\clearpage
\newpage


\begin{thebibliography}{10}

\bibitem{Weart2003}
S.~Weart, {\it Discovery of Global Warming\/} (Harvard University Press, 2003).

\bibitem{Fleagle80}
R.~Fleagle, J.~Businger, {\it An Introduction to Atmospheric Physics\/},
  vol.~25 of {\it International Geophysics Series\/} (Academic Press, Orlando,
  1980), second edn.

\bibitem{Houghton91}
J.~Houghton, {\it The physics of atmospheres\/} (Cambridge University Press,
  Cambridge, U.K., 1991).

\bibitem{Peixoto92}
J.~Peixoto, A.~Oort, {\it Physics of Climate\/} (AIP, N.Y., 1992).

\bibitem{Hartmann94}
D.~Hartmann, {\it Global Physical Climatology\/} (Academic Press, San Diego,
  USA, 1994).

\bibitem{Lacis2010}
A.~Lacis, G.~Schmidt, D.~Rind, R.~Ruedy, {\it Science\/} {\bf 330}, 353 (2010).

\bibitem{Pierrehumbert2011}
R.~Pierrehumbert, {\it Physics Today\/} pp. 33--38 (2011).

\bibitem{Arrhenius1896}
S.~Arrhenius, {\it Philosophical Magazine and Journal of Science\/} {\bf 41},
  237 (1896).

\bibitem{Hulburt1931}
E.~O. Hulburt, {\it Phys. Rev.\/} {\bf 38}, 1876 (1931).

\bibitem{Charney79}
J.~Charney, {\it et~al.\/}, Ad hoc study group on carbon dioxide and climate,
  {\it Ad hoc study group\/}, Climate Research Board, National Research
  Council, USA (1979).

\bibitem{Schneider74}
S.~Schneider, R.~Dickinson, {\it Rev. Geophys.\/} {\bf 12}, 447 (1974).

\bibitem{North75}
G.~North, {\it Journal of the Atmospheric Sciences\/} {\bf 32}, 2033 (1975).

\bibitem{Wang80}
W.-C. Wang, P.~Stone, {\it J. Atmos. Sci.\/} {\bf 37}, 545 (1980).

\bibitem{Anderegg2010}
W.~Anderegg, J.~Prall, J.~Harold, S.~Schneider, {\it PNAS\/} {\bf 107}, 12107
  (2010).

\bibitem{Oreskes2004}
N.~Oreskes, {\it Science\/} {\bf 306} (2004).

\bibitem{IPCC-AR4}
S.~Solomon, {\it et~al.\/}, eds., {\it Climate Change: The Physical Science
  Basis. Contribution of Working Group I to the Fourth Assessment Report of the
  Intergovernmental Panel on Climate Change\/} (Cambridge University Press,
  United Kingdom and New York, NY, USA, 2007).

\bibitem{OreskerConway2008}
N.~Oreskes, E.~Conway, {\it Agnotology: The Making and Unmaking of Ignorance\/}
  (Stanford University Press, 2008), chap. Challenging Knowledge.

\bibitem{ProctorSchiebinger2008}
R.~Proctor, L.~Schiebinger, eds., {\it Agnotology: The Making and Unmaking of
  Ignorance\/} (Stanford University Press, 2008).

\bibitem{Trenberth2004}
K.~Trenberth, D.~P. Stepaniak, {\it Quarterly Journal of the Royal Met.
  Society\/} {\bf 130}, 2677–2701 (2004).

\bibitem{Keeling76}
C.~Keeling, {\it et~al.\/}, {\it Tellus\/} {\bf 6}, 538 (1976).

\bibitem{Trenberth2011}
K.~Trenberth, {\it Climate Research\/} {\bf 47}, 123 (2011).

\bibitem{ERAINT}
A.~Simmons, S.~Uppala, D.~Dee, S.~Kobayashi, Era-interim: New ecmwf reanalysis
  products from 1989 onwards, ECMWF Newsletter (2007).

\bibitem{LiebmannSmith2006}
Liebmann, Smith, {\it Bull. Amer. Meteor. Soc.\/} {\bf 77}, 1275 (2006).

\bibitem{Foster2011}
G.~Foster, S.~Rahmstorf, {\it Environmental Research Letters\/} {\bf 6}, 044022
  (2011).

\bibitem{TheCopenhagenDiagnosis2009}
I.~Allison, {\it et~al.\/}, The copenhagen diagnosis, 2009: Updating the world
  on the latest climate science, {\it Tech. rep.\/}, The University of New
  South Wales Climate Change Research Centre (CCRC), Sydney, Australia (2009).

\bibitem{Santer2003b}
B.~Santer, {\it et~al.\/}, {\it Journal of Geophysical Research\/} {\bf 108},
  4002 (2003).

\bibitem{Vecchi2007}
G.~Vecchi, B.~Soden, {\it Journal of Climate\/} {\bf 20}, 4316 (2007).

\bibitem{Seidel2008}
D.~Seidel, Q.~Fu, W.~Randel, T.~Reichler, {\it Nature geoscience\/} {\bf 1}, 21
  (2008).

\bibitem{Giorgi-et-al-2011}
F.~giorgi, {\it et~al.\/}, {\it Journal of Climate\/} {\bf 24}, 5309 (2011).

\bibitem{Wang2002}
P.-H. Wang, P.~Minnis, B.~Wielicki, T.~Wong, L.~Vann, {\it Geophys. Res.
  Lett.\/} {\bf 29}, 1397 (2002).

\bibitem{Hines2000}
K.~Hines, D.~Bromwich, G.~Marshall, {\it Journal of Climate\/} {\bf 13}, 3940
  (2000).

\bibitem{Bengtsson2004}
L.~Bengtsson, S.~Hagemann, K.~Hodges, {\it Journal of Geophysical Research\/}
  {\bf 109} (2004).

\end{thebibliography}

\end{document}